\documentclass[conference]{IEEEtran}
\IEEEoverridecommandlockouts    

\usepackage{multirow}
\usepackage{lipsum}
\usepackage{amsmath}
\usepackage{amssymb}
\usepackage[ruled,vlined]{algorithm2e}
\usepackage{graphicx}
\usepackage{booktabs}
\usepackage{bbding}
\usepackage{float}
\graphicspath{{./Figures/}}

\title{\LARGE \bf From Sensing to Decision: A Generic Architecture for Freight Signal Priority Systems}

\author{Ziyan Zhang, Xuanpeng Zhao, Chuheng Wei, Ronald William Snyder, Changxin Wan, Peng Hao,~\IEEEmembership{Member,~IEEE}\\
Kanok Boriboonsomsin,~\IEEEmembership{Member,~IEEE}, Guoyuan~Wu,~\IEEEmembership{Senior~Member,~IEEE}
\thanks{Ziyan Zhang, Xuanpeng Zhao, Chuheng Wei, Ronald William Snyder, Changxin Wan, Kanok Boriboonsomsin, Peng Hao and Guoyuan Wu are with the College of Engineering, Center for Environmental Research and Technology, University of California at Riverside, Riverside, CA, 92507.}
\thanks{$^{*}$Corresponding author. e-mail: ziyan.zhang@email.ucr.edu}
\thanks{\textbf{Accepted at ITSC 2026. Final version to appear in IEEE Xplore.}}
}

\begin{document}
	
	\maketitle
	\thispagestyle{empty}
	\pagestyle{empty}
	

	\begin{abstract}
    Freight Signal Priority (FSP) systems have emerged as a promising strategy to enhance freight mobility and reduce corridor delays in urban networks. While extensive research has focused on priority control algorithms and operational performance evaluation, comparatively limited attention has been devoted to the architectural design of sensing processes that shape reliable priority decisions. In practice, uncertainties in vehicle detection, communication, and estimated time of arrival (ETA) may propagate within the sensing-to-decision process, affecting priority timing and downstream signal performance.
    
    This paper presents a systematic review of FSP systems from a sensing-to-decision perspective. We propose a generic two-layer architecture consisting of a sensing-to-decision layer and a control execution layer. The sensing-to-decision layer transforms sensing inputs into priority decisions, while the control execution layer implements approved actions within traffic controllers. Within this architecture, we systematically compare major sensing modalities, including loop detectors, vision sensors, and V2I, across dimensions such as classification capability, state estimation accuracy, latency, and information richness. We further examine representative FSP systems to analyze how modality-specific characteristics and uncertainties influence ETA computation, priority triggering, and decision reliability.
    
    By linking sensing design to decision outcomes, this review identifies key deployment challenges and research gaps in reliability-aware sensing-to-decision design. Ultimately, this work provides a conceptual foundation for developing scalable and robust FSP systems that explicitly account for sensing imperfections rather than assuming idealized inputs.
	\end{abstract}
	
	\section{Introduction}
	\label{sec:introduction}
    Urban traffic signal control has evolved significantly over the past decades, transitioning from fixed-time operations to adaptive and predictive systems capable of responding to real-time traffic conditions \cite{feng2015real}. Early adaptive frameworks such as Real-time Hierarchical Optimized Distributed Effective System (RHODES) introduced hierarchical optimization mechanisms that leveraged short-term vehicle arrival predictions and network-level coordination to improve operational efficiency \cite{mirchandani2001real}. With the emergence of sensing technologies, more advanced adaptive control strategies have been developed, enabling dynamic signal adjustments based on vehicle-level information and real-time communication between infrastructure and road users.

    Building upon adaptive signal control, Signal Priority (SP) systems represent a specialized class of traffic control strategies designed to provide preferential treatment to designated vehicle classes. Unlike conventional signal control that aims to optimize overall traffic performance, SP systems intentionally adjust signal timing, through green extension, early green, or phase insertion, to reduce delay for priority vehicles such as buses, emergency vehicles, and freight trucks. While SP has demonstrated significant reductions in delay and improvements in operational efficiency, its performance critically depends on the accuracy and timeliness of vehicle detection and communication.

    Freight Signal Priority (FSP) has emerged as an increasingly important application of SP, motivated by the disproportionate environmental and operational impacts of heavy-duty freight vehicles. Although freight trucks constitute a small percentage of total registered vehicles, they contribute substantially to fuel consumption and emissions. FSP systems aim to improve freight mobility and reduce stop-and-go operations by accounting for truck-specific characteristics such as longer acceleration times, higher fuel consumption during idling, and grade sensitivity. Recent studies have shown that FSP can reduce freight delay and fuel consumption, particularly under congested or near-capacity conditions \cite{kari2014eco}. However, these benefits are highly sensitive to the reliability of vehicle identification, estimated time of arrival (ETA), and communication latency.

    Despite growing interest in FSP algorithms and performance evaluation, comparatively limited attention has been devoted to the sensing-to-decision processes that shape reliable priority decisions. Most existing studies focus on control logic, optimization models, or eco-driving strategies, typically assuming accurate perception and timely data exchange \cite{zhao2019co, guo2021eco}. Although some FSP implementations incorporate both sensing and control components, their evaluations primarily emphasize control performance rather than examining how sensing uncertainty influences downstream decisions.
    In addition, existing review papers on signal priority predominantly concentrate on priority control strategies and coordination mechanisms, with limited discussion of sensing-layer design and its role in system reliability. In practice, however, uncertainties in vehicle detection, roadside occlusion, and V2I communication latency can significantly distort priority timing and effectiveness. Errors in state estimation or ETA computation may propagate within the sensing-to-decision process, leading to false triggers, missed requests, or unintended impacts on cross-street traffic.
        
    To address this gap, this paper provides a systematic review of FSP system design from a sensing-to-decision perspective. We first present a generic two-layer architecture that organizes FSP operations into a sensing-to-decision layer and a control execution layer, comprising five functional components: sensing, edge processing, edge-level priority decision-making, control interfaces, and signal control. Within this structure, we categorize major sensing modalities and compare their characteristics relevant to priority decision reliability. Furthermore, we examine representative FSP systems and analyze them from the perspectives of sensing modality and their uncertainty impact, deployment scenario, end-to-end sensing-to-control integration, cost, system complexity and deployment maturity. Finally, we discuss practical deployment considerations and research gaps identified in prior studies.
    
    The remainder of this paper is organized as follows. Section II introduces the system architecture. Section III compares sensing modalities. Section IV analyzes representative FSP systems. Section V discusses deployment challenges and future research directions.
    
    \begin{figure*}[h]
        \centering
        \includegraphics[width=0.95\textwidth]{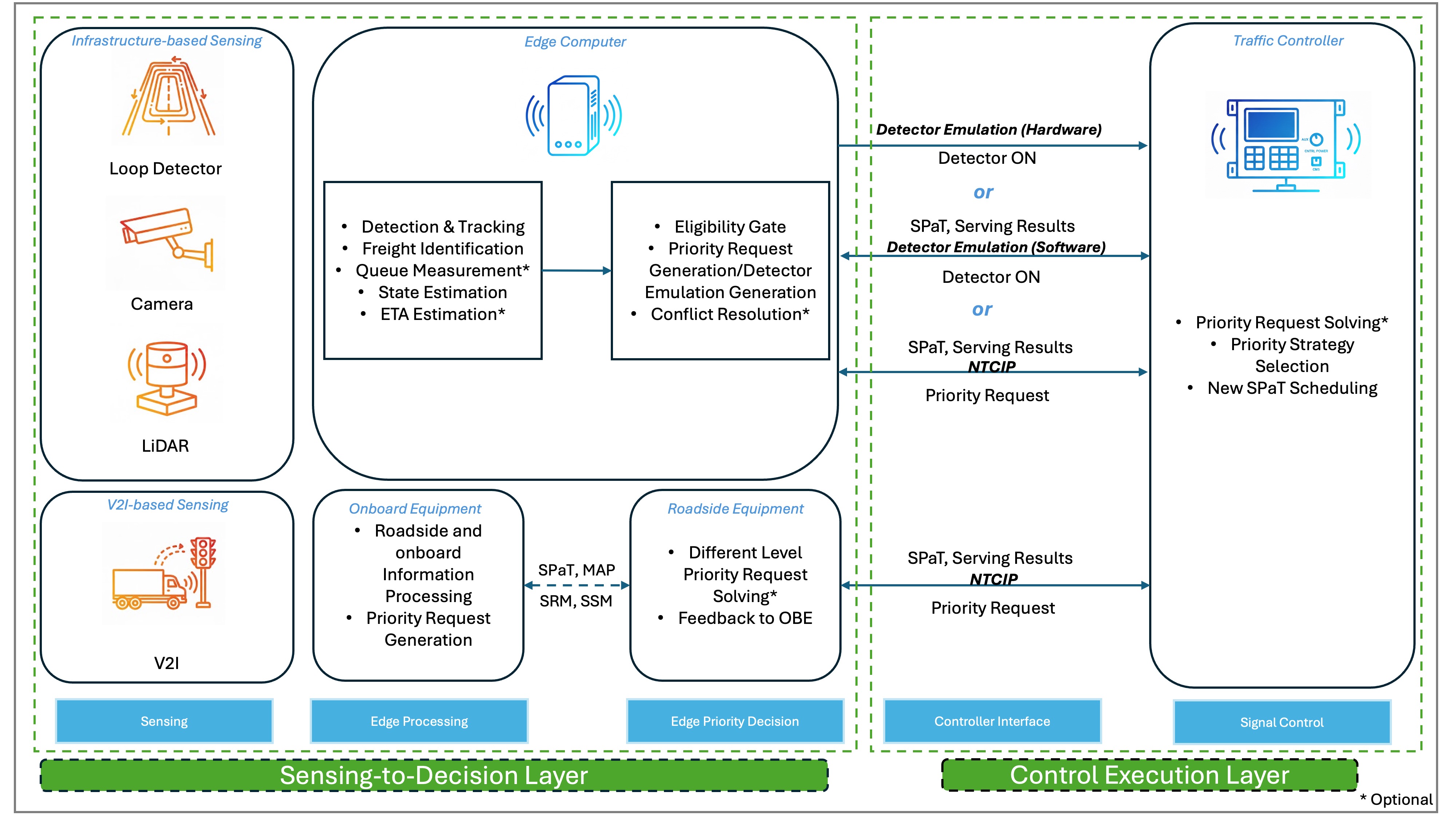}  \caption{Freight Signal Priority System Architecture}
        \label{fig:architecture}
    \end{figure*}


    The main contributions of this paper are summarized as follows:
    \begin{itemize}
    
    \item \textbf{FSP-Centric Sensing Perspective}: We provide the first systematic review of sensing-to-decision design in FSP systems, emphasizing its foundational role in shaping reliable priority decisions.
    
    \item \textbf{Generic Two-Layer FSP Architecture}: We propose a generic two-layer FSP architecture that explicitly distinguishes between the sensing-to-decision layer and the control execution layer, clarifying how sensing inputs are transformed into actionable priority decisions.
    
    \item \textbf{FSP-Oriented Modality Comparison}: We systematically compare sensing modalities based on their impacts on ETA computation, priority triggering, and decision reliability, supported by representative FSP system analyses.
    
    \item \textbf{Reliability and Deployment Insights}: We analyze how sensing uncertainties propagate within the sensing-to-decision process and identify key real-world deployment challenges and research gaps toward reliability-aware FSP system design.
    
    \end{itemize}
	\section{Freight Signal Priority System Architecture}
	\label{sec:FSP Architecture}

    In this section, we introduce a systematic architecture for the FSP system mainly from the perspective of sensing-to-decision. The architecture focuses on an isolated intersection without considering coordinated corridor operations.
    
    We conceptualize the FSP system as consisting of two primary layers: (1) the sensing-to-decision layer and (2) the control execution layer. The sensing-to-decision layer transforms raw sensing inputs into structured priority decisions. The control execution layer implements approved priority requests within the traffic controller.
    
    Within this two-layer structure, the system can be decomposed into five functional components: sensing, edge processing, edge priority decision-making, control interfaces, and signal control. Specifically, sensing, edge processing, and edge priority decision-making together constitute the sensing-to-decision layer, while control interfaces and signal control form the control execution layer.
    
    This review primarily focuses on the sensing-to-decision layer, as it determines how sensing uncertainty propagates into priority decision reliability. Control execution mechanisms are discussed only to clarify downstream impacts.
    
    The FSP architecture can further be categorized into centralized and distributed priority structures, following general Transit Signal Priority (TSP) system designs. In centralized architectures, priority requests are generated at a traffic management center. In distributed architectures, priority decisions are generated locally based on intersection-level sensing results. This paper focuses on distributed architectures, as most existing FSP implementations operate at the local intersection level.
    
    The sensing components considered in this study include loop detectors, cameras, LiDAR, and V2I communication, as illustrated in Fig.~\ref{fig:architecture}. These modalities represent the primary sensing approaches adopted in prior FSP studies.
    In this paper, V2I is treated as a standardized, communication-enabled sensing modality supported by open protocols (e.g., SAE J2735). In contrast, proprietary fleet-based technologies such as Automatic Vehicle Identification (AVI) and Automatic Vehicle Location (AVL) are excluded, as they depend on agency-specific deployments and lack scalability for decentralized freight operations.

    Finally, we distinguish two primary sensing-to-control pipelines based on sensing origin: infrastructure-based sensing and V2I-based sensing. These pipelines differ fundamentally in data acquisition mechanisms, communication dependencies, and uncertainty characteristics, and are therefore analyzed separately in the following sections.

    \subsection{Infrastructure-based Sensing-to-Control Process}
    For infrastructure-based sensing methods, including loop detectors, cameras, and LiDAR, sensors collect vehicle information, and an edge computer processes the data to classify trucks, assign vehicle IDs, estimate vehicle states, and measure queue length when applicable. Based on the processed information, priority decisions are made according to different priority-triggering strategies.
    \subsubsection{Detector Emulation (Hardware)} If hardware-based detector emulation is used, after eligibility filtering, a detector emulation command is generated and sent to the traffic controller through a traditional relay card, mimicking a trigger from a roadway detector.
    \subsubsection{Detector Emulation (Software)} Similarly, if software-based detector emulation is deployed, the edge computer directly sends a mimicked trigger to the traffic controller through dedicated software or the National Transportation Communications for Intelligent Transportation System Protocol (NTCIP). Meanwhile, this method can obtain Signal Phase and Timing (SPaT) information and service status through communication with the traffic controller.
    \subsubsection{NTCIP-Based Request} With the NTCIP-based method, the edge computer generates a priority request containing the vehicle ID, vehicle class, requested phase, estimated time of arrival (ETA), and estimated time of departure (ETD), utilizing SPaT information from the traffic controller. The edge computer then determines the priority decision based on a first-come, first-served rule before sending the request to the traffic controller.
    
    In the final step, the traffic controller determines the appropriate priority strategy (e.g., phase call, green extension, early green, green truncation, phase insertion, or phase skipping) and updates the SPaT accordingly to provide priority to approaching trucks.
    
    \subsection{V2I-based Sensing-to-Control Process}
    For V2I sensing, once a truck enters the communication range, the onboard equipment (OBE) begins receiving MapData (MAP) and SPaT information from the roadside equipment (RSE). Combined with onboard information, the OBE generates a priority request including vehicle ID, vehicle class, requested phase, ETA, ETD, and priority level, and transmits it through a Signal Request Message (SRM). The RSE receives priority requests from multiple trucks and makes a priority decision by considering priority weights or optimizing based on cost. Alternatively, the RSE may simply forward the SRMs to the traffic controller for further processing and execution. After processing an SRM and making a decision, the RSE sends a Signal Status Message (SSM) to the OBE for confirmation. The traffic controller then executes the approved priority action through appropriate phase adjustment strategies.
    
    In the following sections, we focus on the sensing-to-decision process of the FSP system. Control interfaces and specific priority strategies are not discussed in detail in this paper.

	\section{Sensing Modalities Comparison}
	\label{sec:Detection-to-Decision}
    
    \begin{table*}[t]
    \caption{Comparison of Sensing Modalities}
    \label{tab:modality_comparison}
    \centering
    \scalebox{0.87}{
    \begin{tabular}{lccccccc}
    \toprule
    \textbf{Modality} & \textbf{Working Time} & \textbf{Working Weather} & \textbf{Classification Ability} & \textbf{State Estimation Ability} & \textbf{Priority Level Access} & \textbf{Latency} &  \textbf{Detection Range} \\
    \midrule
    Loop Detector & Day \& Night & All Weather & Medium & Medium & No & Lowest & Long-range \\
    Camera & Day & Good Weather &High  & Low & No & Low & Mid-range \\
    LiDAR & Day \& Night & Most Weather & High  & High & No & Low & Mid-range/Long-range \\
    V2I & Day \& Night & All Weather & Highest & Highest & Yes & Medium & Longest-range \\
    \bottomrule
    \end{tabular}
    }
    \end{table*}
    
    Comparing the sensing modalities discussed above, each approach exhibits distinct characteristics that may influence downstream priority decision-making. The analysis presented here considers individual sensing modalities in isolation, without accounting for multimodal fusion. Latency is defined as the combined effect of detection latency and communication latency.
    \begin{itemize}
    \item \textbf{Loop Detector}: It operates continuously under all weather conditions. It can provide basic truck/non-truck classification, but it is difficult to differentiate buses from trucks. Speed estimation is point-based and does not support continuous updates. Vehicle IDs cannot be tracked. Different priority levels for multiple trucks cannot be assigned due to limited information. It has the lowest detection and communication latency. The detection range is predetermined based on assumptions such as average speed and braking distance.
    \item \textbf{Camera}: It operates effectively during daytime and under favorable weather conditions. It can provide truck/non-truck classification, although misclassification may occur. Speed and acceleration estimates derived from vision may be less accurate, leading to ETA errors. Reliable tracking is required to maintain stable vehicle IDs and avoid duplicate priority requests. Different priority levels for multiple trucks cannot be assigned due to limited information. Detection latency is low (depending on computing performance), and communication latency is negligible. The detection range is moderate; distant objects with low resolution may be misclassified or undetected.
    \item \textbf{LiDAR}: It operates continuously in most weather conditions, except during heavy rain, dense fog, or heavy snow. It can provide truck/non-truck classification, although misclassification may still occur. Speed and acceleration estimates are generally accurate, resulting in more reliable ETA estimation. Reliable tracking is required to maintain stable vehicle IDs. Different priority levels for multiple trucks cannot be assigned due to limited information. Detection latency is low but typically higher than that of camera-based systems (depending on computing performance), and communication latency is negligible. The detection range is moderate to long, depending on the LiDAR specifications.
    \item \textbf{V2I}: It operates continuously under all weather conditions. It can provide rich priority request information, including accurate vehicle class, speed, acceleration, stable vehicle ID, and priority level. However, communication latency and packet loss may occur during transmission. It typically offers a longer detection range, allowing more time for priority request processing.
    \end{itemize}
    As summarized in Table~\ref{tab:modality_comparison}, V2I offers the most comprehensive information set among the compared modalities, although its performance is constrained by communication latency and reliability. In practical deployments, however, the selection of sensing modality depends on operational requirements, infrastructure conditions, and budget constraints.
    From an FSP system perspective, representative implementations adopting different sensing modalities are examined in the following section to provide further architectural and deployment insights.
	
    \section{Representative FSP Systems}
    \label{sec:sensing modalities}
    
    In this section, Table~\ref{tab:fsp_systems} categorizes representative FSP systems according to sensing modality, deployment scenario, cost, system complexity, maturity, and their coverage of the sensing-to-decision and control execution layers within the proposed FSP architecture. 
    We compare and analyze these systems to gain insights into how different sensing modalities shape priority decision-making and influence downstream control performance. By examining how sensing-to-decision designs are implemented across modalities, we highlight how architectural choices at this layer affect uncertainty propagation, decision reliability, and real-world deployment effectiveness.
    
    \begin{table*}[t]
    \caption{Taxonomy of Representative FSP Systems}
    \label{tab:fsp_systems}
    \centering
    \scalebox{0.95}{
    \begin{tabular}{lcccccccc}
    \toprule
    \textbf{System} & \textbf{Modality} & \textbf{Year} & \textbf{Scenario} &  \textbf{Sensing-to-Decision} & \textbf{Control Execution} &  \textbf{Cost} & \textbf{Complexity} & \textbf{Maturity} \\
    \midrule
    Sunkari et al. \cite{sunkari2000reducing} & Loop Detector & 2000 & Real World &\checkmark  & \checkmark & Low & Low & High \\
    Mahmud \cite{mahmud2014evaluation} & Loop Detector & 2014 & Simulation &\checkmark  & \checkmark & Low & Low & High \\
    Ardalan \cite{ardalan2020development, kaisar2020evaluation} & Loop Detector & 2020 & Simulation        &\checkmark & \checkmark & Low & Low & High \\
    Belhassine et al. \cite{belhassine2022signal} & Loop Detector & 2022 & Simulation    &\checkmark & \checkmark & Low & Low & High \\
    Chowdhury et al. \cite{chowdhury2023operational} & Loop Detector & 2023 & Simulation        &\checkmark & \checkmark & Low & Low & High \\
    Akkeh et al. \cite{akkeh2024impact} & Loop Detector & 2024 & Simulation  &\checkmark & \checkmark & Low & Low & High \\
    Saunier et al. \cite{Ecole18974} & Camera & 2009 & Real World & $\sqrt{}\mkern-9mu{\smallsetminus}$  & $\times$ & Low & Low & Low \\
    Zhang et al. \cite{zhang2026multi}& Camera \& LiDAR & 2026 & Real World &$\sqrt{}\mkern-9mu{\smallsetminus}$ & $\times$ & Medium & Medium & Low \\
    Ahn et al. \cite{ahn2015multi} & V2I & 2015 & Sim \& Real &\checkmark & \checkmark & High & High & Medium \\
    Park et al. \cite{park2019environmental} & V2I & 2019 & Simulation &\checkmark & \checkmark & Low & High & Medium \\
    Talukder et al. \cite{talukder2022analytical} & V2I & 2022 & Real World          & \checkmark        & \checkmark    & High  & High & Medium \\
    Cvijovic et al. \cite{cvijovic2022signal} & V2I & 2022 & Simulation &\checkmark  & \checkmark    & Low  & High & Medium \\
    Das et al. \cite{das2023priority} & V2I & 2023 & Simulation   &\checkmark    & \checkmark    & Low  & High & Medium \\
    Zhao et al. \cite{ zhao2019co} &  V2I  & 2019 & Simulation  & $\times$ & \checkmark        & Low      & High  & Medium \\
    \bottomrule
    \end{tabular}
    }
    \end{table*}

    \subsection{\textbf{Sensing Modalities and Their Uncertainty Impact}}
    While existing studies primarily evaluate sensing modalities based on detection accuracy and feasibility, their influence on downstream priority control performance is often implicitly treated. In this subsection, we explicitly analyze how modality-specific sensing uncertainties propagate along the sensing-to-decision pipeline, through state estimation, ETA computation, priority request generation, and phase actuation, and ultimately influence signal control stability and effectiveness.
    \subsubsection{Loop Detector}

    As shown in Table \ref{tab:fsp_systems}, loop-detector-based FSP systems were among the earliest approaches explored due to their technological maturity and low cost. However, loop detectors provide point-based vehicle detection without continuous tracking or direct vehicle identification. Consequently, ETA estimation relies on assumed speed profiles and predetermined detector placement.
    
    Previous studies \cite{mahmud2014evaluation, ardalan2020development, belhassine2022signal} determine detector distance based on clearance time, speed limits, or estimated truck speeds. Some researchers adopt the Minimum Stopping Sight Distance (MSSD) defined by AASHTO \cite{hancock2013policy}, while others incorporate driver reaction time, detector processing time, and truck length \cite{sunkari2000reducing}. Akkeh et al. \cite{akkeh2024impact} adopted the conventional detector placement methodology for FSP design. However, due to limited storage length in the left-turn lane, the detector could not be installed at the theoretically designed location. Instead, it was placed at the upstream end of the left-turn lane, reflecting a practical constraint that alters the intended sensing configuration.
    
    From an uncertainty propagation perspective, loop detectors introduce spatial and kinematic assumptions into the control pipeline. Any deviation between assumed and actual truck speeds directly affects ETA estimation. Such ETA bias shifts the timing of priority requests, which may lead to premature green extensions, missed early-green opportunities, or unnecessary disruption of cross-street phases. Therefore, sensing uncertainty in detector placement and speed assumptions propagates into phase timing decisions and intersection efficiency.
    
    \subsubsection{Vision Sensors}

    Relatively few studies have integrated vision-based sensing (camera or LiDAR) into FSP systems. Saunier et al. \cite{Ecole18974} proposed a prototype sensing system achieving 78\%–95\% recall with a false alarm rate below 0.5\%; although the system covers detection function, most of the functions in sensing-to-decision layer and control execution layer are not integrated. Only vehicle classification was provided without ETA computation, limiting its applicability for real-time priority triggering. More recently, \cite{zhang2026multi} proposed a multi-modal sensing system for FSP at both intersection and midblock locations, though full integration remains limited.
    
    Vision-based systems introduce edge processing-layer uncertainties, including misclassification, tracking instability, and state estimation noise. These uncertainties propagate through inaccurate ETA computation or duplicate/missed vehicle identification. False positives may trigger unnecessary priority calls and phase extensions, whereas false negatives eliminate intended priority benefits. Unstable tracking may generate repeated requests for the same vehicle, disrupting phase sequencing and reducing signal timing stability.
    
    Thus, edge processing-layer uncertainty directly influences priority service reliability, phase balance, and overall traffic performance. Unlike loop detectors, vision-based systems reduce spatial assumptions but introduce stochastic perception errors that affect control stability.

    \subsubsection{V2I}

    With the advancement of Intelligent Transportation System (ITS), FSP sensing has gradually shifted toward V2I-based approaches, particularly within the Multi-modal Intelligent Traffic Signal System (MMITSS) framework. Talukder et al. \cite{talukder2022analytical} demonstrated improved ETA accuracy in coordinated corridors. Ahn et al. \cite{ahn2015multi} reported up to 20\% delay reduction for connected trucks in field deployment. Subsequent studies \cite{das2023priority, park2019environmental} further evaluated V2I-enabled multi-modal priority systems.
    
    Although V2I provides richer vehicle information, including speed, acceleration, and unique identifiers, it introduces communication-layer uncertainties such as latency, packet loss, and penetration variability. Temporal uncertainty shifts the arrival time of priority requests relative to feasible actuation windows. Dropped or delayed SRMs may cause missed green extensions or incorrect sequencing of competing priority calls.
    
    Therefore, communication-layer uncertainty propagates into control performance by reducing timing precision and increasing service variability. In coordinated corridors, even small latency fluctuations may accumulate across intersections, potentially destabilizing progression quality and diminishing corridor-level performance gains.

    Across sensing modalities, uncertainty sources differ in nature: spatial assumptions for loop detectors, perception instability for vision sensors, and communication variability for V2I. However, their impacts converge at the control execution layer through distorted ETA estimation, mistimed priority requests, and altered phase actuation decisions. This convergence highlights the necessity of reliability-aware priority control strategies that explicitly account for sensing-to-detection imperfections, rather than assuming deterministic and error-free inputs.

    \subsection{\textbf{Deployment Scenario}}

    From a scenario perspective, most FSP systems are developed and evaluated primarily in simulation environments. This approach is understandable given the high cost and complexity of real-world infrastructure deployment. Some simulation-based studies integrate both sensing-to-decision and control execution layers; however, sensing and communication processes are often simplified or idealized, limiting the representation of real-world operational variability.
        
    Hybrid evaluation approaches that combine simulation and field deployment \cite{ahn2015multi} provide a more realistic assessment of the FSP system. Such integration can further evolve into digital twin architectures, where real-time sensing data are synchronized with predictive models to enable closed-loop performance evaluation and reliability-aware optimization.

    \subsection{\textbf{End-to-End Sensing-to-Control Integration}}

    Existing vision-based FSP systems have generally not integrated sensing-to-decision processes with priority control evaluation. In addition, Table~\ref{tab:fsp_systems} includes a representative study that focused solely on control design while abstracting the sensing-to-decision process \cite{zhao2019co}. Many simulation-based control studies assume ideal V2I communication without realistic communication constraints or priority request generation mechanisms when evaluating FSP performance \cite{kari2014eco, suthaputchakun2015novel}.
        
    Such perfect-information assumptions effectively decouple the sensing-to-decision layer from the control execution layer and prevent assessment of how sensing uncertainty propagates into priority timing and phase actuation decisions. Consequently, the interaction between the sensing-to-decision layer and control logic remains underexplored, highlighting a critical research gap in end-to-end FSP system evaluation.

    \subsection{\textbf{Cost}}
    From a cost perspective, loop detectors, cameras, and LiDAR generally incur lower infrastructure-level expenses, whereas V2I systems require additional investment for communication infrastructure and onboard equipment (OBE) deployment across vehicle fleets. Although this results in higher initial costs, large-scale OEM pre-installation of standardized OBEs could reduce per-vehicle expenses over time.
    Therefore, cost trade-offs should be evaluated in conjunction with sensing richness, integration complexity, scalability, and long-term operational benefits rather than device-level expenses alone.

    \subsection{\textbf{System Complexity and Deployment Maturity}}
    From a deployment perspective, loop-detector-based FSP systems exhibit the simplest architecture and the highest maturity, benefiting from seamless integration with conventional traffic controllers. However, their limited sensing capability constrains flexibility and scalability.
    V2I-based systems demonstrate moderate maturity through implementations such as MMITSS and provide richer vehicle information with improved timing precision. Nevertheless, they introduce substantial system complexity, including OBEs, RSUs, wireless communication infrastructure, and multi-layer message processing, which increases integration and maintenance burdens.
    Vision-based FSP systems fall between these two approaches in terms of complexity. They eliminate the need for fleet-wide OBEs but require advanced perception algorithms and high-performance edge computing. Despite strong sensing potential, challenges related to occlusion, control integration, and large-scale validation remain, resulting in relatively low deployment maturity.
    
    Overall, a trade-off exists between architectural simplicity and sensing richness. Higher maturity often comes with limited capability, whereas advanced sensing solutions increase integration complexity and deployment difficulty.
    
	\section{Deployment Challenges and Research Gaps}
	\label{sec:challenges}
    \subsection{Deployment Challenges}
    \subsubsection{\textbf{Vision Occlusion}}
    When cameras or LiDAR are used for sensing, occlusion may occur due to terrain-related or object-related factors. For example, if an intersection or corridor includes road curvature within the predetermined detection range of the sensor for FSP purposes, approaching trucks may not be detected in a timely manner, thereby affecting priority decisions.
    In addition, trucks may be occluded by roadside vegetation of similar height. In multi-truck scenarios, a following truck may be occluded by a leading truck, and a truck in the inner lane may be occluded by a truck in the outer lane. Such occlusions may result in missed triggers or delayed priority service for the affected vehicle.
    For challenging terrain conditions, researchers should select appropriate sensing modalities or adjust deployment locations, such as moving sensing units from the main intersection to midblock locations, as demonstrated in \cite{zhang2026multi}. To mitigate object occlusion, sensor placement and projection angles should be carefully optimized to maximize coverage. Simulation-based evaluation using virtual sensors in platforms such as CARLA \cite{dosovitskiy2017carla} or specialized toolboxes \cite{zheng2025inspe} can support deployment planning. Furthermore, installing multiple sensors and applying sensor fusion to extend coverage is another effective strategy \cite{zimmer2023tumtraf}.

    \subsubsection{\textbf{Time Synchronization}}
    Time synchronization across OBEs, RSEs, edge computers, and traffic controllers is a critical deployment challenge. Clock drift and inconsistent timestamping may introduce ETA estimation errors, misinterpretation of SPaT information, and degraded priority effectiveness, particularly in coordinated corridors. As reported in \cite{zhang2026multi}, synchronization issues were observed between different sites connected via wireless 5 GHz AirFiber links.
    Potential solutions include GPS-based time synchronization, Precision Time Protocol (PTP), and Network Time Protocol (NTP), with decreasing synchronization accuracy in that order.

    \subsubsection{\textbf{Communication Latency and Packet Loss}}
    V2I-based FSP relies on timely and reliable message exchange among OBEs, RSEs, and the signal control interface. In real-world deployments, communication latency and packet loss may arise from channel congestion, non-line-of-sight (NLOS) conditions, and hardware or software bottlenecks. At busy intersections, frequent message broadcasts and multiple concurrent priority requests (e.g., SRMs) increase medium access contention, resulting in stochastic delay, jitter, and occasional packet loss. Urban canyon effects, large-vehicle blockage, and suboptimal antenna placement may further degrade link quality and cause burst losses, particularly near the communication range boundary.
    Latency and packet loss directly affect the timing precision and service reliability of priority operations. If a priority request arrives late, the controller may miss the feasible actuation window for early green or green extension. If requests are dropped, the priority call may not be issued at all. Similarly, delayed or lost downstream broadcasts (e.g., SPaT or SSM) may mislead vehicle-side decision-making and reduce the effectiveness of priority strategies. Therefore, deployments should incorporate reliability-aware communication design. In addition, field calibration of RSU placement, antenna height and orientation, and channel load monitoring is necessary to maintain stable performance under varying traffic demand.

    \subsection{Research Gaps}
    \subsubsection{\textbf{Perfect Assumption for Sensing in Simulation}}
    Sensing uncertainties remain among the most overlooked factors affecting the real-world performance of FSP systems. Simulation tools are widely used to evaluate FSP control schemes prior to deployment; however, sensing and communication processes are often abstracted or idealized, implicitly assuming accurate detection and timely data exchange. As a result, even under the same control strategy, variations in sensing-to-decision design and associated uncertainties may alter ETA estimation and priority request timing, leading to divergent performance outcomes. For example, assuming perfect V2I connectivity without latency or stable detection without false positives and false negatives may yield overly optimistic conclusions that fail to generalize to real-world conditions. Therefore, uncertainty modeling should be treated as an essential component of simulation-based research rather than a secondary consideration.
    Several approaches can mitigate this gap. For V2I simulation, communication latency can be incorporated into priority triggering time \cite{yelchuru2014aeris}, or statistical delay models can be embedded within communication modules (e.g., VISSIM’s Message Distributor \cite{fellendorf2010microscopic}). More rigorously, co-simulation with network simulators such as ns-3 \cite{henderson2008network} enables dynamic coupling between wireless channel conditions and traffic density.
    For sensor simulation, coupling traffic simulators with scenario simulators that incorporate realistic sensor models (e.g., SUMO \cite{lopez2018microscopic} with CARLA) offers a practical solution. Cyber-mobility frameworks \cite{bai2022cyber}, which integrate real-world sensing data with simulation, further improve deployment realism.
    
   \subsubsection{\textbf{Accuracy Requirements for Sensor Detection}}
    Roadside detection has attracted increasing attention in recent years~\cite{wei2025hierarchical}. For FSP systems, detection serves as the foundational component of the sensing-to-decision process, directly affecting ETA estimation accuracy and priority triggering reliability. Therefore, high detection accuracy and stable state estimation are essential to ensure dependable priority performance.
    Long-range LiDAR is particularly suitable for truck detection in FSP applications due to its extended coverage and precise state estimation capability. However, as noted in \cite{zhang2026multi}, the lack of large-scale roadside detection datasets for long-range LiDAR, especially solid-state LiDAR, limits model generalization and detection performance. This data scarcity constrains achievable detection accuracy and highlights a research gap in developing comprehensive long-range roadside datasets tailored to FSP scenarios.
    In addition, sensor fusion offers a promising pathway to improve detection robustness and reduce uncertainty. However, effective fusion requires multi-modal roadside datasets specifically designed for freight-priority applications.
    
    \subsubsection{\textbf{Other Possible Sensors for FSP}}
    Although RADAR sensors have been widely deployed for traffic signal actuation and vehicle detection, their application in FSP systems remains largely unexplored. RADAR provides accurate speed measurement and long-range detection under all weather conditions at relatively low cost~\cite{wei2025integrating}. In addition, object length and Doppler features may support truck classification.
    Emerging 4D LiDAR technologies provide $(x, y, z)$ coordinates and radial velocity for each point, enabling more accurate speed estimation and ETA calculation~\cite{urazghildiiev2007vehicle}. However, the cost of 4D LiDAR remains significantly higher than that of conventional 3D LiDAR.
    
    \subsubsection{\textbf{FSP on Left Turns}}
    Most existing FSP systems focus on through movements. However, limited storage length and restricted green time in left-turn phases require targeted research on left-turn FSP operations. As noted earlier, insufficient storage length may prevent loop detectors from being installed at the theoretically designed distance, leading to ETA estimation bias and timing errors. Akkeh et al. \cite{akkeh2024impact} identified the lack of detector placement guidelines for left-turn FSP scenarios.
    Advanced sensing modalities such as cameras, LiDAR, and V2I can mitigate this spatial constraint. Continuous tracking and trajectory inference for turning from vision or LiDAR, as well as turning-related information transmitted via V2I (e.g., BSM-based turn signal status), reduce reliance on fixed detector placement and improve robustness under geometric limitations~\cite{frossard2019deepsignals}.

    \subsubsection{\textbf{FSP for Platooning Trucks}}
    Chowdhury et al. \cite{chowdhury2023operational} reported that, under existing traffic control systems, truck platooning may increase travel time and the number of stops for both passenger vehicles and trucks. High penetration rates of truck platooning can further exacerbate delay, even with signal priority.
    This highlights the need for FSP strategies specifically designed for platooning operations, considering platoon length and headway characteristics. From a sensing perspective, since platooning systems already employ vehicle-to-vehicle (V2V) technologies, onboard equipment can also support V2I communication. Moreover, given the severe occlusion effects within tightly spaced platoons, V2I sensing may provide more reliable detection compared with infrastructure-only sensing.

    \subsubsection{\textbf{Digital Twin for FSP}}
    As discussed previously, integration of simulation and real-world FSP operations remains limited. Most existing systems are evaluated either through standalone simulation or isolated field deployment, without real-time cyber–physical synchronization. Although such systems demonstrate operational feasibility, they do not constitute true digital twin architectures, as they lack continuous state reconstruction, predictive modeling, and closed-loop feedback integration.
    A digital twin framework would enable dynamic state estimation and network-level performance optimization, supporting reliability-aware FSP decision-making. Compared with platforms such as MMITSS, a digital twin system can reconstruct real-world scenarios in a virtual environment and continuously evaluate and optimize FSP decisions based on traffic state prediction, with closed-loop error correction between the virtual and physical systems~\cite{ali2022digital}.
    
    \subsubsection{\textbf{LLM/VLM Applications for FSP}}
    Large Language Models (LLMs) and Vision-Language Models (VLMs) have recently been explored in Transportation Systems Management and Operations (TSMO) to enhance traffic system monitoring and decision support~\cite{hassan2025large}. In the context of FSP, LLM/VLM technologies may provide new capabilities for sensing interpretation, contextual reasoning, and coordination support.
    For example, a roadside management agent could leverage VLM-based scene understanding to interpret real-time traffic conditions on main and side streets, while an onboard agent integrates vehicle-level information. Such systems could assist in priority coordination and generate explainable textual reports or operational guidance for drivers and traffic management personnel, rather than directly replacing conventional control logic.
    
	\section{Conclusion}
	\label{sec:conclusion}

This paper synthesizes FSP system design from a sensing-to-decision perspective and proposes a generic two-layer architecture comprising a sensing-to-decision layer and a control execution layer. Our analysis demonstrates how sensing uncertainty, arising from detector placement assumptions, perception instability, and communication variability, propagates through ETA estimation and priority request timing, ultimately shaping signal control performance.

We show that end-to-end integration between sensing-to-decision and control execution remains limited, particularly in vision-based implementations and simulation-driven studies that assume ideal conditions. From a system design perspective, a clear trade-off emerges: richer sensing improves information quality but increases overall system integration complexity, whereas simpler designs achieve higher maturity with reduced flexibility.

Addressing deployment challenges such as occlusion, synchronization, and communication reliability, together with closing identified research gaps, will be essential for next-generation FSP systems. Ultimately, robust FSP performance depends on explicitly modeling and managing uncertainty within the sensing-to-decision layer to enable scalable and deployment-ready implementations.
	
	\bibliographystyle{IEEEtran}
	\bibliography{root} 
	
\end{document}